# Hybrid AC/DC Transmission Expansion Planning Considering HVAC to HVDC Conversion Under Renewable Penetration

M. Moradi-Sepahvand, and T. Amraee, *Senior Member, IEEE*

*Abstract*—In this paper, a dynamic (i.e. multi-year) hybrid model is presented for Transmission Expansion Planning (TEP) utilizing the High Voltage Alternating Current (HVAC) and multiterminal Voltage Sourced Converter (VSC)-based High Voltage Direct Current (HVDC) alternatives. In addition to new HVAC and HVDC lines, the possibility of converting existing HVAC transmission lines to HVDC lines is considered in the proposed model. High shares of renewable resources are integrated into the proposed hybrid AC/DC TEP model. Due to the intermittency of renewable resources, the planning of large-scale Energy Storage (ES) devices is considered. In order to accurately estimate the total TEP costs and hence capturing the scenarios of load and renewable generation uncertainty, using a clustering approach, each year of the planning horizon is replaced with four representative days. The proposed model is formulated as a Mixed-Integer Linear Programming (MILP) problem. Using Benders Decomposition (BD) algorithm, the proposed model is decomposed into a Master investment problem to handle the decision variables, and Sub-problems to check the feasibility of master problem solution and optimize the operation and ES investment cost. Three test systems are used as case studies to demonstrate the effectiveness of the proposed hybrid AC/DC TEP model.

## NOMENCLATURE

*Indices & Sets:*

| | |
|---|---|
| $t, \Omega_T, T$ | Index, Set, and Total number of planning years. |
| $d, \Omega_D, D$ | Index, Set, and Total number of representative days in each year. |
| $h, \Omega_H, H$ | Index, Set, and Total number of hours in each day. |
| $i, \Omega_B$ | Index and Set of all buses. |
| $j, \Omega_G$ | Index and Set of Generator buses. |
| $l$ | Index of all HVAC and all HVDC lines. |
| $l'$ | Index of single and double circuit candidate lines for AC to DC conversion as sub-index of $l$. |
| $\Omega_{NL}^{AC}, \Omega_{NC}^{AC}, \Omega_{EL}$ | Sets of all candidate HVAC lines, new HVAC lines in new corridors as a subset of $\Omega_{NL}^{AC}$, existing HVAC lines. |
| $\Omega_{all}^{DC}, \Omega_{con}^{DC}$ | Sets of all HVDC candidate lines, candidate lines for AC/DC conversion as a subset of $\Omega_{all}^{DC}$. |
| $v, \Omega_V$ | Index and Set of Voltage Source Converters. |
| $\Omega_{SB}$ | Set of candidate buses for Energy Storage devices. |
| $c, \Omega_C$ | Index and Set of the allowable candidates in the same corridor. |
| $p, \Omega_P, P$ | Index, Set, and Total number of linear segments of the generation cost function. |

*Parameters:*

| | |
|---|---|
| $r$ | Interest rate. |
| LT | Lifetime of equipment (year). |
| $IC_l$ | Investment cost of new HVAC and HVDC lines ($Million/Km). |
| $RWC_l$ | Right of Way cost for new HVAC and HVDC lines in new corridors ($ Million/Km). |
| $LL_l, ASC_l$ | Line length in new HVAC and HVDC corridors (Km), HVAC new Substation Cost ($ Million). |
| $Cg_p^j, VC_v$ | Cost of power generation of conventional units in each segment for each unit ($/MW), VSCs cost ($ Million/MVA). |
| $\{\bullet\}^{max}/\{\bullet\}^{min}$ | Maximum/minimum limits of bounded variables. |
| $A_l^i, K_l^i$ | The $(i, l)$ element of directional Connectivity matrixes of existing and new HVAC lines with busses. |
| $\alpha, \beta$ | Constant factors of the VSC power loss function. |
| $C_e, C_P$ | Construction cost of energy capacity ($/MWh) and power capacity ($/MW) of ES devices. |
| $\eta_c, \eta_d$ | Charging and discharging efficiency of ESs. |
| $WF_h^d, LF_h^d$ | Hourly Wind and Load forecast factor. |
| $Kcb_v^i, Kc_l^v$ | The $(i, v)$ and $(v, l)$ elements of connectivity matrixes of VSCs with buses and all HVDC lines with VSCs. |
| $Ld_i^{PK}, LG$ | Peak load of bus $i$ (MW), Load Growth factor. |
| $\rho_d$ | Probability (i.e. normalized time duration) of the occurrence of each representative day $d$. |
| $\xi$ | The reserve cost factor as a percentage of the cost of power generation. |
| $Kcn_{l'}^l$ | Discriminant matrix of single and double circuit existing lines selected for conversion to HVDC. |
| $M, \Psi, B$ | Big-M, the Base power of the system (MVA), per unit susceptance of all HVAC lines. |

*Variables:*

| | |
|---|---|
| $Ya_{l,c}^t, Yd_{l,c}^t$ | Binary variables of candidate HVAC and HVDC lines at year $t$ and corridor $c$ (equals 1 if the candidate line is constructed and 0 otherwise). |
| $Yc_l^t, Ycc_l^t$ | Binary variables of HVAC single and double circuit lines selected for conversion at year $t$. |
| $Ye_l^t, Pe_{l,d,h}^t$ | Binary variable of existing HVAC lines at year $t$, Power flow of existing HVAC lines at year $t$, day $d$ and hour $h$ (MW). |

M. Moradi-Sepahvand and T. Amraee are with the Department of Electrical Engineering, K.N. Toosi University of Technology, Tehran, Iran. (e-mails: mojtaba.moradi@email.kntu.ac.ir, amraee@kntu.ac.ir ).

| | |
|---|---|
| $I_{j,d,h}^t$, $U_{i,d,h}^t$ | On/off state of generating unit *j* at year *t*, day *d*, and hour *h*, Charging/discharging state of ES in bus *i* at year *t*, day *d*, and hour *h*. |
| $Ss_i^t$, $Cs_i^t$ | Total energy (MWh) and power (MW) capacity of ES *i* in year *t*. |
| $P_{j,d,h}^t$, $Pg_{j,d,h,p}^t$ | Power output of conventional unit *j*, at year *t*, day *d*, and hour *h*, Power generation of segment *p* of unit *j*, at year *t*, day *d*, and hour *h* (MW) |
| $Pw_{j,d,h}^t$, $R_{j,d,h}^t$ | Power output of wind farm *j*, at year *t*, day *d*, and hour *h* (MW), Reserve of unit *j* at year *t*, day *d*, and hour *h* (MW). |
| $Ps_{i,d,h}^t$, $Es_{i,d,h}^t$ | Power output (MW), and stored energy (MWh) of ES in bus *i* at year *t*, day *d*, and hour *h*. |
| $Pd_{i,d,h}^t$, $Pc_{i,d,h}^t$ | Discharging and charging power of ES in bus *i* at year *t*, day *d*, and hour *h* (MW). |
| $Pl_{l,c,d,h}^t$ | Power flow of new constructed HVAC lines, in corridor *c*, at year *t*, day *d*, and hour *h* (MW). |
| $Pvc_{v,l,d,h}^t$ | Power capacity of installed VSC *v* for HVDC line *l*, at year *t*, day *d*, and hour *h* (MW). |
| $\theta_{i,d,h}^t$ | Voltage angle of bus *i* at year *t*, day *d*, and hour *h*. |

*Compact Representation:*

| | |
|---|---|
| **Y** | Vector of binary decision variables. |
| **S** | Vector of ES energy and power capacity. |
| **P** | Vector of positive continuous operational variables. |
| **Q** | Vector of free continuous variables. |

## I. INTRODUCTION

### A. Background and Literature Review

HVDC transmission lines have been recognized as a techno-economic solution for bulk power transfer, especially across the long-distance corridors. Reduction of total planning cost, power losses, transmission congestion relief, and providing direct control of power flows are important features of HVDC lines [1]. HVDC lines can be used for transferring renewable generation from remote sites to the demand centers [2-3]. The hybrid AC/DC Transmission Expansion Planning (TEP) model refers to the network expansion model in which both HVAC and HVDC options are considered [1-2].

In the presence of renewable energy resources, the energy storage (ES) devices are utilized for removing the possible congestion, minimizing the renewable curtailment and deferring transmission expansion [4]. In addition to HVDC lines and ES devices, the conversion of existing HVAC lines to HVDC ones, is generally seen as an effective way to increase the power transfer capability of the existing transmission system [5-6]. In [7], a technical comparison is conducted between HVDC and hybrid AC/DC systems. By converting HVAC to HVDC lines, the total transmitted power in an existing corridor can be increased up to 3.5 times [8].

The conventional TEP using HVAC lines has been well addressed in previous research works [9-12]. The benefits of ES devices in conventional TEP studies (i.e. the full HVAC TEP models) are investigated in [13, 14]. In [13], the delays in transmission expansion and the degradation in storage capacity are also taken into account under different renewable and load growth scenarios. In [14], the co-planning of conventional HVAC-based network and the ES devices is investigated with minimizing the investment cost and facilitating the power system operation under high renewable shares.

Transmission expansion models can be either full HVDC-based or hybrid AC/DC. In [15], a market-based long-term HVDC-based TEP model is presented for meshed VSC-HVDC grids that connect regional markets. The model is used to investigate development of the offshore grid in the North Sea with maximizing congestion revenue and minimizing investment costs of the HVDC grid. In [16], the long-term planning of the meshed HVDC-based grid is addressed and three different converter models including nonlinear, second order convex relaxation, and linear approximation models are compared. The investment costs of HVDC system and the generation costs are optimized.

The focus of the present paper is to develop a hybrid AC/DC TEP model in the presence of renewable resources, considering ES devices, and HVAC to HVDC conversion. A stochastic AC/DC TEP problem for supplying load forecasts, minimizing investment costs, and optimizing market operations is discussed in [1]. In [2], a TEP model is proposed for integrating renewable resources considering AC/DC transmission alternatives. In [17], an MILP model is proposed for solving the multi-stage TEP problem, including security constraints using HVAC and HVDC alternatives. Also, in [17], the authors indicate that better expansion plans can be achieved using HVDC lines in the expansion process.

In [18], a long term TEP model is proposed using both AC and DC transmission lines. Based on the shortest path algorithm, the proposed MIP model is iteratively solved. In [19], a hybrid AC/DC stepwise TEP model is proposed for the Spanish and French transmission systems. The impact of construction delays and availability of multi-terminal HVDC lines is investigated. The authors in [19], conclude that for the long-distance and high power transmission links, mostly HVDC technology is preferred. In [20], a hybrid AC/DC TEP model is proposed. The uncertainty of remote renewable resources and load demand is considered and the proposed model is solved using the binary differential evolution algorithm. In [21], a single stage or static robust formulation is presented for ES and AC/DC transmission systems co-planning. Authors in [21] conclude that ES devices can alleviate transmission congestion and results in a better transmission plan. In [22] and [23], the impacts of HVDC lines in optimal power flow of hybrid AC/DC networks are addressed. Due to the high cost of new Right of Way (RoW) and higher power transfer capability, the conversion of HVAC to HVDC lines seems to be a favorable investment decision. In previously proposed TEP problems, the conversion of HVAC lines to HVDC has not been addressed. Also, the consideration of renewable and load scenarios in combination with ES devices for accurate modeling of hybrid AC/DC TEP is still a major gap that is addressed in this research.

### B. Contributions

This paper presents a hybrid HVAC/HVDC TEP model in the presence of renewable resources, and HVAC lines, HVDC lines, HVAC to HVDC conversion, and ES devices are considered as decision variables. The main contributions of this paper are summarized as follows.
1) A multi-year HVAC/HVDC TEP model is developed which allows the efficient utilization of modern tools such

as HVDC options and ES devices in transmission expansion planning. Since the consideration of ES devices in hybrid AC/DC TEP models needs to replace each year with some representative days, a clustering approach [24], is utilized to combine the load and wind scenarios and extract four representative days. The extracted representative days capture the interday and intraday variations of load and renewable scenarios, and then the daily cycles of ES devices are easily handled in the proposed hybrid AC/DC model.

2) In addition to the individual HVAC lines, multiterminal VSC-based HVDC lines, and ES devices, the conversion of existing HVAC lines to HVDC lines is formulated and considered as a powerful transmission expansion tool. This tool is designed for both single and double circuit HVAC lines. In the case of a double circuit HVAC line, the conversion of either one or both AC systems is properly modeled.

3) An efficient methodology based on the Benders Decomposition (BD) algorithm is developed to handle the complexity of the proposed multi-year AC/DC TEP problem. The investment costs of transmission lines and charging/discharging state of ES devices are handled in the Master Problem (MP), while the feasibility of MP solution, the investment cost of required ES devices and optimization of the operation cost are managed in Dual Sub-Problem (DSP).

The rest of this paper is organized as follows. In section II, the overall structure of the proposed model is presented. The detailed formulation of the proposed hybrid HVDC/HVAC model is given in Section III. The simulation results of the proposed method over the test cases are discussed in Section IV. Finally, the significant findings are given in Section V.

## II. STRUCTURE OF THE PROPOSED MODEL

The general structure of the proposed model is illustrated in Fig. 1. The proposed model is formulated as an MILP problem and is solved using the BD algorithm for efficient handling of investment and operation costs. Indeed, the proposed hybrid AC/DC TEP model is a very complicated problem with different decision variables. The cost of HVAC, HVDC, HVAC to HVDC conversion, ES devices and the operation costs of power generation should be optimized with considering a vast range of technical constraints along a multi-year horizon. Therefore, a BD-based method is developed to handle all these parts. The BD technique allows optimizing the binary and continuous variables in different master and sub-problems. As it is illustrated in Fig. 1, at first, the required inputs are defined, and then in the BD algorithm, two levels of optimization, i.e. MP and DSP, are handled. The MP minimizes the total investment cost of new constructed HVAC lines, HVDC lines and AC to DC conversion. Also, the charging/discharging states of ES devices are determined in MP. The total investment cost of ES devices and the total operation cost including generation and reserve costs of conventional units are minimized in DSP. Based on decision variables obtained from MP, the generated power of conventional and renewable units, power flows of lines, power flows across the multi-terminal VSCs and capacity of ES devices are determined using DSP. After solving DSP, if the solution is not unbounded and the deference between Lower Bound (LB), generated by MP, and the Upper Bound (UB), determined by DSP, is less than a pre-defined threshold, the expansion plan is obtained and the algorithm is finished; otherwise, the linear BD optimality/feasibility cuts are constructed and sent to the MP. In Fig. 1, the numbers of main equations are reported, and detailed explanations are presented in later sections.

## III. FORMULATIONS

In this section, the complete formulations of the proposed model including the general and BD formulation, are presented.

### A. General Formulation

The objective function and the related constraints of the proposed hybrid model are given below.

*1) Objective Function*

In the majority of expansion planning studies, the investment costs of new devices and operation costs are considered as the objective function (e.g. [1], [13] and [17]). The operation costs can be optimized using a cost-based mechanism in which the generation costs over the planning horizon is minimized. The generation costs can be optimized using a market-based mechanism in which the social welfare (i.e. the benefits of consumers and generation companies) is maximized. In this paper, the cost-based mechanism is used to optimize the generation costs. According to (1), the general objective function of the proposed multi-year hybrid AC/DC TEP model minimizes two main costs, i.e. Total Investment Cost (TIC) and Total Operation Cost (TOC). The total planning cost (i.e. Z) represents the discounted present values of the investment cost of all new transmission lines (i.e. A, B), VSCs (i.e. C), ES installation (i.e. D), and the hourly operation cost of generating units. The discounted present values of the investment and operation costs at each year are assumed at the beginning and end of that year, respectively.

$$Min\ Z = \sum_{t \in \Omega_T} \left[ \left( \frac{1}{(1+r)^{t-1}} \right) TIC_t \right] + \left[ \left( \frac{1}{(1+r)^t} \right) TOC_t \right] \quad (1)$$

In (1), the TIC consists of four terms as expressed by (2):

$$TIC_t = A_t + B_t + C_t + D_t \quad (2)$$

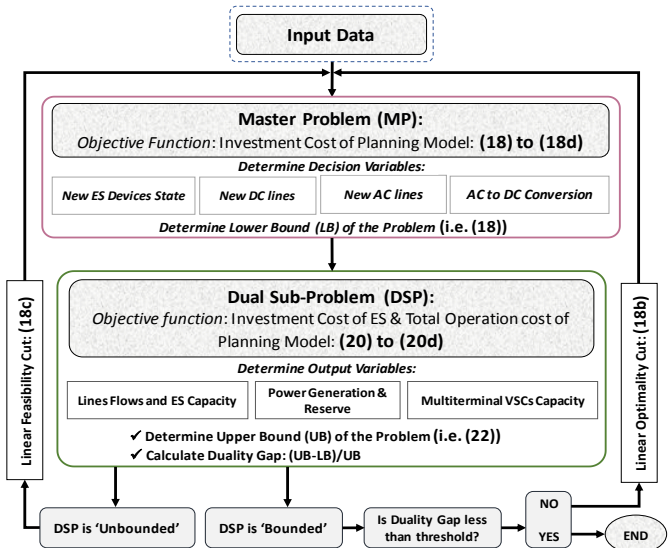

Fig. 1. The General framework of the proposed hybrid AC/DC TEP model

In (2), the discounted annual cost of all new HVAC lines in existing and new corridors for each year of planning is computed as given in (2a). It is assumed that in existing HVAC single circuit corridors, it is possible to add one new HVAC or HVDC line to the existing towers without any additional RoW cost. In new corridors, it is assumed that new constructed lines are all double circuit, and the cost of AC substation are considered only for the first new HVAC line (i.e. $c=1$).

$$A_t = K_{Lac} \left[ \sum_{l \in \Omega_{NL}^{AC}} IC_l \left( \sum_{c \in \Omega_C} (Ya_{l,c}^t - Ya_{l,c}^{t-1}) \right) \right.$$
$$+ \sum_{l \in \Omega_{NL}^{AC}} \left[ ((RWC_l).(LL_l)). \sum_{c \in \Omega_C} (Ya_{l,c}^t - Ya_{l,c}^{t-1}) \right]$$
$$\left. + \sum_{l \in \Omega_{NC,c=1}^{AC}} \left[ (Ya_{l,c}^t - Ya_{l,c}^{t-1}).(ASC_l) \right] \right] \quad (2a)$$

The discounted annual cost of all new HVDC lines in existing and new corridors for each year is expressed as given in (2b).

$$B_t = K_{Ldc} \left[ \sum_{l,l \in \Omega_{all}^{DC}} IC_l \left[ \left( \sum_{c \in \Omega_C} (Yd_{l,c}^t - Yd_{l,c}^{t-1}) \right) \right. \right.$$
$$\left. + (Yc_l^t - Yc_l^{t-1}) + \left( \frac{1}{2} (Ycc_l^t - Ycc_l^{t-1}) \right) \right]$$
$$\left. + \sum_{l \in \Omega_{all}^{DC}} \left[ (RWC_l).(LL_l). \sum_{c \in \Omega_C} (Yd_{l,c}^t - Yd_{l,c}^{t-1}) \right] \right] \quad (2b)$$

The discounted annual VSCs investment cost for all HVDC lines and converted lines is represented by (2c).

$$C_t = K_V \left[ \sum_{v \in \Omega_V} \sum_{l \in \Omega_{all}^{DC}} Kc_l^v.P_l^{max}.VC_v \left[ \left( \sum_{c \in \Omega_C} (Yd_{l,c}^t - Yd_{l,c}^{t-1}) \right) + (Yc_l^t - Yc_l^{t-1}) + \frac{1}{2} (Ycc_l^t - Ycc_l^{t-1}) \right] \right] \quad (2c)$$

Finally, the discounted annual investment cost of ES devices considering the rated energy and power of each ES is considered as follows.

$$D_t = \sum_{i \in \Omega_{SB}} \left[ K_S (Ss_i^t - Ss_i^{t-1}) + K_P (Cs_i^t - Cs_i^{t-1}) \right] \quad (2d)$$

In addition to the investment costs, the TEP configuration should minimize the TOC, including the generation and reserve costs of conventional units as given by (3):

$$TOC_t = \sum_{d \in \Omega_D} \left[ 365.\rho_d \sum_{h \in \Omega_H} \left[ \sum_{j \in \Omega_G} \left[ Cg_{p=1}^j.(P_j^{min}.I_{j,d,h}^t + \xi.R_{j,d,h}^t) \right] \right. \right.$$
$$\left. \left. + \sum_{p \in \Omega_P} \left[ Cg_p^j.Pg_{j,d,h,p}^t \right] \right] \right] \quad (3)$$

It is noted that the operation cost of wind farms and ES devices is negligible. The $K_{Lac}$, $K_{Ldc}$, $K_V$, $K_S$ and $K_P$ are Capital Recovery Factor (CRF) of new HVAC and HVDC lines, VSCs and ESs, respectively, defined by (4) to (6). It is noted that the CRFs given in (4) is written in a compact form.

$$K_{Lac}, K_{Ldc}, K_V = \frac{r(1+r)^{LT_{Line,VSC}}}{(1+r)^{LT_{Line,VSC}} - 1} \; (\$/year) \quad (4)$$

$$K_S = C_e \frac{r(1+r)^{LT_{ES}}}{(1+r)^{LT_{ES}} - 1} \; (\$/MWh.year) \quad (5)$$

$$K_P = C_P \frac{r(1+r)^{LT_{ES}}}{(1+r)^{LT_{ES}} - 1} \; (\$/MW.year) \quad (6)$$

*2) Constraints of Generating Units*

The focus of the present paper is to present a hybrid AC/DC TEP while the generation cost is considered as a major part of total planning cost. For a better approximation of total generation cost, the economic loading of generators and their power generation limits are added to the proposed formulation. Also, the cost function of conventional generating units is linearized. However, other technical limits of generating units, such as Minimum Up Time, Minimum Down Time, Ramp Rate, Start-up and Shut Down limits are not considered. The constraints of all conventional and renewable units are represented by (7) to (8). According to (7a), the minimum and maximum generation capacity of units are satisfied. In (7b), for linearization of the nonlinear cost function of conventional units, the power generation of each unit is expressed as a set of linear segments plus the minimum generated power. In (7c), the limits of power generation in each segment are defined. In (8), the limits of wind generation considering hourly forecast factors for each year are presented. The binary variable $I$ is defined as the hourly on/off state of generating units to determine the economic loading of generators.

$$P_j^{min}.I_{j,d,h}^t \leq P_{j,d,h}^t \leq P_j^{max}.I_{j,d,h}^t \quad \forall j \in \Omega_G, t \in \Omega_T, d \in \Omega_D, h \in \Omega_H \quad (7a)$$

$$P_{j,d,h}^t = P_j^{min}.I_{j,d,h}^t + \sum_{p=1}^P Pg_{j,d,h,p}^t \quad \forall j \in \Omega_G, t \in \Omega_T, d \in \Omega_D, h \in \Omega_H \quad (7b)$$

$$0 \leq Pg_{j,d,h,p}^t \leq (P_j^{max} - P_j^{min}). I_{j,d,h}^t / P$$
$$\forall j \in \Omega_G, t \in \Omega_T, d \in \Omega_D, h \in \Omega_H, p \in \Omega_P \quad (7c)$$

$$0 \leq Pw_{j,d,h}^t \leq (WF_h^d.Pw_j^{max}).I_{j,d,h}^t \quad \forall j \in \Omega_W, t \in \Omega_T, d \in \Omega_D, h \in \Omega_H \quad (8)$$

*3) Reserve constraints*

The constraints given in (9a) to (9c) are defined for modeling the reserve requirement. According to (9c), the total reserve is assumed to be greater than the summation of 5% and 3% of the forecasted wind and load, respectively [13]. The integration of renewable resources needs a new kind of spinning reserve named by flexible ramp reserve to handle the uncertainty of renewable resources. To this end, in this paper, it is assumed that a minimum of 5% of the forecasted wind generation is required to be assigned as flexible ramp reserve.

$$0 \leq R_{j,d,h}^t \leq P_{j,d,h}^t \quad \forall j \in \Omega_G, d \in \Omega_D, t \in \Omega_T, h \in \Omega_H \quad (9a)$$

$$R_{j,d,h}^t + P_{j,d,h}^t \leq P_{j,t}^{max} \quad \forall j \in \Omega_G, t \in \Omega_T, d \in \Omega_D, h \in \Omega_H \quad (9b)$$

$$\sum_{j \in \Omega_B} R_{j,d,h}^t \geq (0.05) \times \sum_{j \in \Omega_W} Pw_{j,t}^{max}$$
$$+ (0.03) \times \sum_{i \in \Omega_B} Ld_i^{PK}.LF_d^h.(1+LG)^t \quad \forall t \in \Omega_T, d \in \Omega_D, h \in \Omega_H \quad (9c)$$

*4) Existing and Candidate HVAC Lines Constraints*

In this paper, it is possible to convert an existing HVAC single/double circuit line to a single/double circuit Bipolar HVDC line. In the case of a double circuit HVAC line, either one or both AC systems can be converted. The constraints given in (10a) and (10b) represent the branch flow and upper/lower flow limits of existing AC lines, considering the possibility of conversion of one circuit of double circuit HVAC lines to HVDC.

$$-M_l.(1 - Ye_l^t) \leq Pe_{l,d,h}^t - \sum_{i \in \Omega_B} \Psi.B_l.A_l^i.\theta_{i,d,h}^t \leq M_l.(1 - Ye_l^t)$$
$$\forall l \in \Omega_{EL}, t \in \Omega_T, d \in \Omega_D, h \in \Omega_H \quad (10a)$$

$$-P_l^{max}.\left[Ye_l^t + \sum_{l'\in\Omega_{con}^{DC}} Kcn_{l'}^l.(\frac{Ycc_{l'}^t}{2})\right] \le Pe_{l,d,h}^t \le$$
$$P_l^{max}.\left[Ye_l^t + \sum_{l'\in\Omega_{con}^{DC}} Kcn_{l'}^l.(\frac{Ycc_{l'}^t}{2})\right] \quad \forall l \in \Omega_{EL}, t \in \Omega_T, d \in \Omega_D, h \in \Omega_H \tag{10b}$$

Details about the selection of big-M can be found in [25]. Based on (11a) and (11b), the branch flow of the new candidate AC lines is limited. The constraint in (11c) guarantees that the constructed lines in a given year remain in the system until the end of the planning horizon.

$$-M_l.(1-Ya_{l,c}^t) \le Pl_{l,c,d,h}^t - \sum_{i\in\Omega_B} \Psi.B_l.K_l^i.\theta_{i,d,h}^t \le M_l.(1-Ya_{l,c}^t) \tag{11a}$$
$$\forall l \in \Omega_{NL}^{AC}, c \in \Omega_C, t \in \Omega_T, d \in \Omega_D, h \in \Omega_H$$
$$-P_l^{max}.Ya_{l,c}^t \le Pl_{l,c,d,h}^t \le P_l^{max}.Ya_{l,c}^t \tag{11b}$$
$$\forall l \in \Omega_{NL}^{AC}, t \in \Omega_T, d \in \Omega_D, h \in \Omega_H$$
$$Ya_{l,c}^t \ge Ya_{l,c}^{t-1} \quad \forall l \in \Omega_{NL}^{AC}, c \in \Omega_C, t \in \Omega_T \tag{11c}$$

### 5) Constraints of HVDC Lines

VSC-based HVDC gives more technical benefits than the LCC-based HVDC and it is now the preferred technology in HVDC transmission systems [26]. VSC-based HVDC technology creates more flexibility and its capability in both active and reactive power is a very important characteristic. The aim of the proposed hybrid AC/DC TEP study is to find the optimal transmission plan regarding long term concerns while taking into account the operational aspects as much as possible. The voltage disturbance mitigation or reactive power control is a major advantage of VSC-HVDC technology, that is not addressed in this study. A multiterminal VSC HVDC line with converter stations is illustrated in Fig. 2. In this part, for modeling HVDC lines, VSC losses are defined as given by (12) [2]:

$$P_{loss}^V = \alpha + \beta Pvc_{ac\_side} + \gamma(Pvc_{ac\_side})^2 \cong \alpha + \beta Pvc_{ac\_side} \tag{12}$$

As illustrated in Fig. 2, only the linear terms of (12) are modeled and the nonlinear term, related AC side transformer and also the lines losses are ignored because the noticeable part of HVDC system power losses is related to the converter. The constant factors in (12) are assumed according to [27]. The coupling constraint between the AC and DC sides of each VSC is expressed by (13a) for all HVDC lines. In (13b), the flow limits of VSCs are defined according to the lines connected to VSCs.

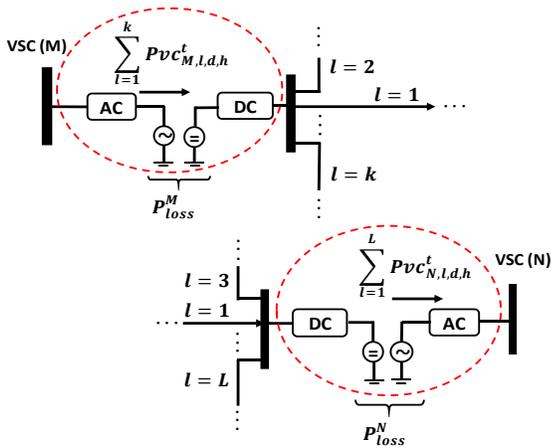

Fig. 2. The structure of multiterminal HVDC VSC stations

$$\sum_{v\in\Omega_V} Kc_l^v\left(1-\beta.\sum_{c\in\Omega_C} Yd_{l,c}^t\right) Pvc_{v,l,d,h}^t = 2\alpha.\sum_{c\in\Omega_C} Yd_{l,c}^t \tag{13a}$$
$$\forall l \in \Omega_{all}^{DC}, t \in \Omega_T, d \in \Omega_D, h \in \Omega_H$$
$$-(Kc_l^v.P_l^{max}.\sum_{c\in\Omega_C} Yd_{l,c}^t) \le Pvc_{v,l,d,h}^t \le (Kc_l^v.P_l^{max}.\sum_{c\in\Omega_C} Yd_{l,c}^t) \tag{13b}$$
$$\forall v \in \Omega_V, l \in \Omega_{all}^{DC}, t \in \Omega_T, d \in \Omega_D, h \in \Omega_H$$
$$Yd_{l,c}^t \ge Yd_{l,c}^{t-1} \quad \forall l \in \Omega_{all}^{DC}, c \in \Omega_C, t \in \Omega_T \tag{13c}$$

### 6) Constraints of HVAC to HVDC Conversion

Four configurations are possible for AC to DC conversion including Monopolar, Bipolar, Tripolar, and Hybrid [26, 28]. In this paper, as shown in Fig. 3, the Bipolar and Hybrid configurations are considered for conversion. Constraints in (14a) and (14b) define the coupling constraint and flow limits of VSCs for converted lines. Based on (14c), if one AC system of a double circuit line is converted to DC, the remained AC circuit is able to transfer half of the maximum capacity of the double circuit AC system. Based on (14d), the conversion state of existing HVAC lines is modeled and (14e) bounds the maximum capacity of each VSC for all its connected HVDC lines.

$$\sum_{v\in\Omega_V} Kc_{l'}^v(1-\beta)Pvc_{v,l',d,h}^t = 2\alpha(Yc_{l'}^t + Ycc_{l'}^t) \tag{14a}$$
$$\forall l' \in \Omega_{con}^{DC}, t \in \Omega_T, d \in \Omega_D, h \in \Omega_H$$
$$-P_{l'}^{max}(Kc_{l'}^{vsc}.\left[Yc_{l'}^t + Ycc_{l'}^t/2\right]) \le Pvc_{v,l',d,h}^t$$
$$\le P_{l'}^{max}(Kc_{l'}^v.\left[Yc_{l'}^t + Ycc_{l'}^t/2\right]) \tag{14b}$$
$$\forall v \in \Omega_V, l' \in \Omega_{con}^{DC}, t \in \Omega_T, d \in \Omega_D, h \in \Omega_H$$
$$-M_l.\left(1-\sum_{l'\in\Omega_{con}^{DC}} Kcn_{l'}^l.Ycc_{l'}^t\right) \le Pe_{l,d,h}^t - \sum_{i\in\Omega_B} S_b.\frac{B_l}{2}.A_l^i.\theta_{i,d,h}^t$$
$$\le M_l.\left(1-\sum_{l'\in\Omega_{con}^{DC}} Kcn_{l'}^l.Ycc_{l'}^t\right) \quad \forall l \in \Omega_{EL}, t \in \Omega_T, d \in \Omega_D, h \in \Omega_H \tag{14c}$$
$$Ye_l^t + \sum_{l'\in\Omega_{con}^{DC}} Kcn_{l'}^l.(Yc_{l'}^t + Ycc_{l'}^t) = 1 \quad \forall l \in \Omega_{EL}, t \in \Omega_T \tag{14d}$$
$$\sum_{l\in\Omega_{all}^{DC}} Pvc_{v,l,d,h}^t \le Pvc_v^{max} \quad \forall v \in \Omega_V, t \in \Omega_T, d \in \Omega_D, h \in \Omega_H \tag{14e}$$
$$Yc_{l'}^t \ge Yc_{l'}^{t-1} \quad \forall l' \in \Omega_{con}^{DC}, t \in \Omega_T \tag{14f}$$
$$Ycc_{l'}^t \ge Ycc_{l'}^{t-1} \quad \forall l' \in \Omega_{con}^{DC}, t \in \Omega_T \tag{14g}$$

### 7) Energy Storage Constraints

In (15a) the power exchange of ESs is defined. Constraints in (15b) and (15f) define the ESs rated power and energy capacity, respectively. Based on (15c) and (15d) the charging and discharging states of ES do not occur at the same time.

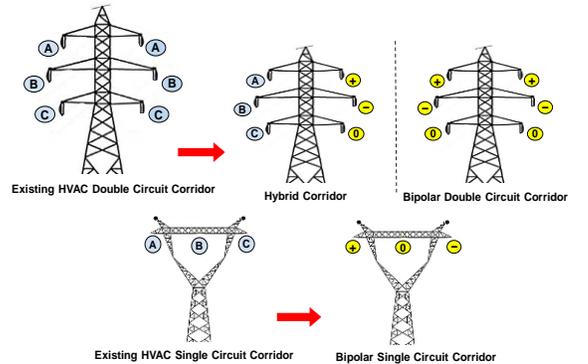

Fig. 3. Configurations for HVAC Single/Double Circuit Lines conversion

According to (15e), the energy level of ES at each hour consists of two terms. Assuming a given hour, the first term presents the energy level of ES at the previous hour. The second term is the energy exchange due to ES charging/discharging. The maximum energy and power capacity of ESs are limited, according to (15g) and (15h).

$$Ps_{i,d,h}^t = Pd_{i,d,h}^t - Pc_{i,d,h}^t \quad \forall i \in \Omega_{SB}, t \in \Omega_T, d \in \Omega_D, h \in \Omega_H \tag{15a}$$

$$0 \leq Pd_{i,d,h}^t \,\&\, Pc_{i,d,h}^t \leq Cs_i^t \quad \forall i \in \Omega_{SB}, t \in \Omega_T, d \in \Omega_D, h \in \Omega_H \tag{15b}$$

$$Pc_{i,d,h}^t \leq Cs^{max}(1 - U_{i,d,h}^t) \quad \forall i \in \Omega_{SB}, t \in \Omega_T, d \in \Omega_D, h \in \Omega_H \tag{15c}$$

$$Pd_{i,d,h}^t \leq Cs^{max} \cdot U_{i,d,h}^t \quad \forall i \in \Omega_{SB}, t \in \Omega_T, d \in \Omega_D, h \in \Omega_H \tag{15d}$$

$$Es_{i,d,h}^t = Es_{i,d,h-1}^t + \left[(\eta_c Pc_{i,d,h}^t) - \left(\frac{1}{\eta_d}Pd_{i,d,h}^t\right)\right] \tag{15e}$$

$$0 \leq Es_{i,d,h}^t \leq Ss_i^t \quad \forall i \in \Omega_{SB}, t \in \Omega_T, d \in \Omega_D, h \in \Omega_H \tag{15f}$$

$$0 \leq Ss_i^t \leq Ss^{max} \quad \forall i \in \Omega_{SB}, t \in \Omega_T \tag{15g}$$

$$0 \leq Cs_i^t \leq Cs^{max} \quad \forall i \in \Omega_{SB}, t \in \Omega_T \tag{15h}$$

$$Ss_i^t \geq Ss_i^{t-1} \quad \forall i \in \Omega_{SB}, t \in \Omega_T \tag{15i}$$

$$Cs_i^t \geq Cs_i^{t-1} \quad \forall i \in \Omega_{SB}, t \in \Omega_T \tag{15j}$$

### 8) Power Balance Constraint

According to (16), the power balance is satisfied at all time intervals. The power balance includes the generation power of conventional and renewable units, demands in buses, power flow of existing and new HVAC lines, new HVDC and converted lines, and power exchange of ESs.

$$P_{i,d,h}^t + Pw_{i,d,h}^t + Ps_{i,d,h}^t - (Ld_i^{PK} \cdot LF_h^d \cdot (1+LG)^t) = \sum_{l \in \Omega_{EL}} A_l^i Pe_{l,d,h}^t + \sum_{l \in \Omega_{NL}^{AC}, c \in \Omega_C} K_l^i Pl_{l,c,d,h}^t + \sum_{l \in \Omega_{all}^{DC}, v \in \Omega_V} Kcb_v^i Pvc_{v,l,d,h}^t \tag{16}$$

$$\forall i \in \Omega_B, t \in \Omega_T, d \in \Omega_D, h \in \Omega_H$$

### B. Solution Technique

The proposed dynamic hybrid TEP is a large-scale MILP problem that needs an efficient solution algorithm. The solution procedure is based on the framework illustrated in Fig. 1. Using BD algorithm, the model is decomposed into an MP to handle the decision variables, and a DSP to check the feasibility of MP solution and optimize the operation and ES investment cost. In order to summarize the formulations of the used BD algorithm, the following Compact Form (CF) is defined.

### 1) Compact Form

The defined CF is presented as follows.

$$\text{Min } I_L^T Y + I_S^T S + O_C^T P \tag{17}$$

s.t.

$$AY = B \tag{17a}$$
$$CY \geq D \tag{17b}$$
$$EP + FQ = G \quad : \sigma \tag{17c}$$
$$H_1 Y + J_1 S + K_1 P + L_1 Q = M \quad : \lambda \tag{17d}$$
$$H_2 Y + J_2 S + K_2 P + L_2 Q \geq N \quad : \mu \tag{17e}$$

$Y \in \{0,1\}, \quad S, P \geq 0, \quad Q: free$
$Y = \{Y_a, Y_e, Y_d, Y_c, Y_{cc}, U, I\}, Q = \{\theta, P_l, P_e, P_s, P_{vc}\}$
$P = \{P, P_w, P_g, P_c, P_d, E_s\}, \quad S = \{S_s, C_s\}$
$\sigma\,\&\,\lambda: free, \quad \mu \geq 0$

The CF is introduced based on the general formulation presented in Part A. The objective function represented by (17) consists of cost function given by (1) to (3) in general formulation. The compact equality constraint in (17a) denotes the constraint given in (14d) and the inequality constraint in (17b) models (11c), (13c), (14g) and (14f). Constraint (17c) represents the constraint of (16). The constraint given in (17d) contains the constraints of (7b), (13a), (14a), (15a) and (15e). Constraint (17e) mimics the rest of constraints of the general formulation (i.e. (7a), (7c), (8), (9a)-(9c), (10a), (10b), (11a), (11b), (13b), (14b), (14c), (14e), (15b)-(15d), (15f)-(15j)). Due to the advantages of Duality formation, such as independency of its search space from integer variables and constructing valid cuts [29], DSP is preferred in this paper. It is noted that $\sigma, \lambda$ and $\mu$ are the dual variables of corresponding constraints. $I_L$ and $I_S$ are the investment cost vectors and $O_C$ is the operation cost vector. Finally, $A, B, C, D, E, F, G, H_1, H_2, J_1, J_2, K_1, K_2, L_1, L_2, M$ and $N$ are relevant matrices.

### 2) Master Problem

According to the proposed CF, the proposed MP is formulated as an Integer Programming (IP) problem as follows:

$$\text{Min } Z_{lower} \tag{18}$$

s.t.

$$Z_{lower} \geq I_L^T Y \tag{18a}$$

$$Z_{lower} \geq I_L^T Y + [M^T \lambda + N^T \mu + G^T \sigma]^{(v)} + \pi^{(v)} Y \tag{18b}$$

$$[M^T \lambda + N^T \mu + G^T \sigma]^{(v)} + \pi^{(v)} Y \leq 0 \tag{18c}$$

$$(17a)\,\&\,(17b) \tag{18d}$$

The objective function given in (18) calculates the Lower Bound (LB) and (18a) denotes investment cost in feasibility condition (i.e. when DSP is unbounded). Constraints given by (18b) and (18c) are the optimality and feasibility cuts, respectively. Constraint (18d) includes the mentioned constraints (17a) and (17b). The index $v$ expresses number of iterations and $\pi$ is the dual variable of (19), which is one of the SP constraints.

$$IY_{sp} = \overline{Y} \quad : \pi$$
$$I: \text{Identity Matrix}, \; \pi: free \tag{19}$$

### 3) Dual Sub-Problem

The DSP is formulated as a Linear Programming (LP) problem by (20) to (20d).

$$\text{Max } M^T \lambda + N^T \mu + G^T \sigma + \overline{Y}^T \pi \tag{20}$$

s.t.

$$J_1^T \lambda + J_2^T \mu \leq I_S \tag{20a}$$
$$K_1^T \lambda + K_2^T \mu + E^T \sigma \leq O_C \tag{20b}$$
$$L_1^T \lambda + L_2^T \mu + F^T \sigma = 0 \tag{20c}$$
$$H_1^T \lambda + H_2^T \mu + \pi \leq 0 \tag{20d}$$

The objective function of the DSP is represented by (20). The corresponding constraints of the DSP are defined via (20a) to (20d). After solving MP (i.e. (18)-(18d)), the decision variables (i.e. $\overline{Y}$) are transferred to the DSP as constants. Afterward, if the DSP has a feasible solution, the optimality cut is constructed and transferred to the MP, otherwise for generating feasibility cut the Modified DSP is solved.

### 4) Modified DSP

Modified DSP (MDSP) is a model for eliminating unbounded conditions in DSP search space and generating feasibility cut [29]. The objective function of MDSP is similar to (20) and its constraints are as (20a) to (20d), except that the Right-Hand-Side (RHS) of these constraints is equal to zero. Additionally, a new auxiliary constraint is considered as given in (21).

$$\sigma \leq 1 \tag{21}$$

It must be noted that the Upper Bound is calculated by (22), which is the summation of the DSP's objective function and the investment cost of transmission lines.

$$UB = M^T\lambda + N^T\mu + G^T\sigma + \overline{Y}^T\pi + I_L^T\overline{Y} \quad (22)$$

## IV. SIMULATION RESULTS

In this section, the performance of the proposed hybrid model is investigated using three test systems including the Garver's 6-bus, IEEE 24-bus and IEEE 118-bus systems. In this paper, it is assumed that the generation expansion plan is known and the TEP model is executed to find the optimal transmission plan over the given horizon. All the decision investments are assumed at the beginning of each year. For example, if it is required to build a new transmission line at year t, it means that the related transmission line should be ready (i.e. in service) at the beginning of that year. The existing network is assumed as the input of the problem. In other words, the state of the problem at the beginning of the planning horizon is similar to the existing network. The time horizon of the planning is 7 years. The LT is considered as 50 years for all HVAC, HVDC and converted lines and VSCs. Also, for ES devises the LT is assumed 10 years. The cost of ES devices is assumed to be 500 $/kw and 25 $/kwh [13], and both the charging/discharging efficiency are considered as 0.9. The investment cost of HVAC, HVDC lines, HVAC substations, VSCs and new RoW are assumed as 1 and 0.96 $10^6$$/km, 2.55 $10^6$$, 0.201 $10^6$$/MW and 0.04 $10^6$$/km, respectively [30]. The investment cost of HVAC to HVDC line conversion is considered 0.13 $10^6$$/km based on [5]. Besides, it is assumed that the existing towers and insulations remain unchanged and therefore, the AC to DC conversion can increase the rating power of lines up to 80%. The maximum number of lines in each corridor is assumed to be 5 for all HVAC, HVDC and converted lines. The yearly load growth and interest rate are both considered equal to 5%. All MILP models in this paper, are solved by CPLEX solver in GAMS using a PC with Intel Core i7, 4.2 GHz 7700 CPU, and 32 GB DDR4 RAM.

The simulation results are presented for two different test cases. In part A, the result of extracting representative days for each year of the planning is presented. In part B, the results of the proposed method over two test cases are discussed in detail.

### A. Representative Days

In multi-year planning, due to computational complexity, it is not necessary to consider a whole year (i.e. 365 days or 8760 hours points) of operation. In this paper, the hierarchical clustering approach [24], is used to extract four representative days, considering the correlation between load and wind variations. In the hierarchical clustering approach, the possibility of merging the pair of clusters is checked based on the minimum variance criteria. The initial clusters are formed using the squared Euclidean distance between elements. The centroid of each cluster is then defined. Centroid refers to the mean of elements in each cluster. Afterward, a dissimilarity index is defined between each pair of clusters and is used to merge the similar clusters. The dissimilarity index is defined as the minimum variance between clusters. The historical data of load and wind power are extracted from [31] and [32],

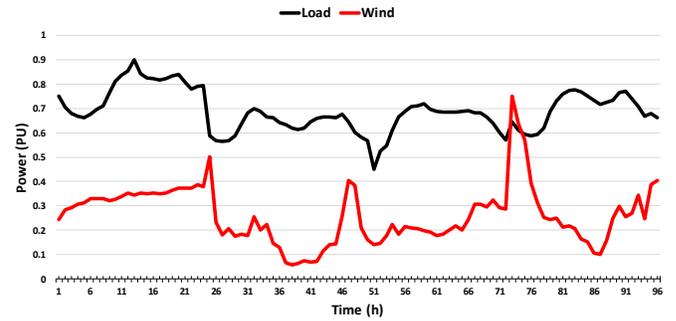

Fig. 4. The load and wind generation of representative days

respectively. Fig. 4, shows the profiles of load and wind in selected representative days.

### B. Case Studies

Three case studies are utilized to verify the effectiveness of the proposed hybrid AC/DC TEP model.

#### 1) Garver's 6-bus Test System

The modified Garver's 6-bus test system consists of five buses and six existing lines. Also, two buses (i.e. buses 6 and 7), can be connected to the system by four candidate HVAC and four candidate HVDC lines. Bus 7 contains a 500 MW wind farm. In this system, there is no double circuit line and five existing single circuit lines are considered as AC to DC conversion candidates. In addition, three HVAC, and three HVDC candidate lines are assumed as expansion candidates. Four buses are considered as candidate locations for ES installation. The complete parameters of all existing and candidate lines and also generating units can be found in [33]. The results of proposed dynamic hybrid TEP for Garver's 6-bus test system are presented with and without considering line losses. To this end, the power loss of existing and new constructed HVAC and HVDC lines, and the power loss of converted HVAC to HVDC lines are considered using the piecewise linear approximation method as proposed in [17].

The four considered schemes are as follows: **scheme 1** is the Dynamic Hybrid TEP (DHTEP) without ES and AC to DC conversion, **scheme 2** is DHTEP considering ES but without AC to DC conversion, **scheme 3** is DHTEP considering AC to DC conversion but without ES and finally, **scheme 4** is DHTEP which considers both ES and AC to DC conversion. Additionally, for all schemes, a new scheme with Considering Line Losses (CLL) is simulated. Indeed, in four main schemes, just the power loss of converter stations is considered, but in their related CLL case, the losses of all lines are also considered. The results of main four schemes and their related CLL cases are reported in Table I. Also, in Fig. 5, the results of scheme 4 are illustrated. The obtained results confirm that the scheme 4 is more economical than other schemes. In this scheme, the total planning cost is 8.296 $10^6$$, 5.792 $10^6$$ and 16.677 $10^6$$ less expensive than schemes 3, 2 and 1, respectively. Comparison of schemes 4, 3 and 2 with 1, confirms the beneficial impacts of ES and HVAC to HVDC line conversion on reducing the total planning cost. According to the last column of Table I, the fourth scheme results in construction of three HVDC lines in corridors 2-6, 4-6, and 5-7 (i.e. 1×(2-6), 1×(4-6), 1×(5-7)), and one HVAC in corridor 2-6, all at the first year of planning. Construction of both HVAC and HVDC lines in corridor 2-6, confirm the importance of hybrid AC/DC TEP.

TABLE I
The complete planning results of the Garver's 6-bus test system

| Scheme | TIC (10⁶$) | | | | TOC (10⁶$) |
|---|---|---|---|---|---|
| | HVAC/DC Lines | ESs | Conversion | Total | 1644.547 |
| 1 | 111.867 | ― | ― | 111.867 | TPC** (10⁶$):<br>Z=1756.414 |
| PD* | HVAC: 1×(3-5) & 1×(2-6) **t=1**, 1×(3-5) **t=2**, 1×(2-6) **t=5**, 1×(4-6) **t=7**<br>HVDC: 1×(4-6), 1×(2-6) & 1×(5-7) **t=1**<br>*CPU Time: 263 Sec.* | | | | |
| CLL*** | 119.520 | ― | ― | 119.520 | 1690.759<br>Z=1810.279 |
| PD | HVAC: 1×(3-5) & 2×(2-6) **t=1**, 1×(3-5) **t=2**<br>HVDC: 1×(2-6), 1×(4-6) & 1×(5-7) **t=1**, 1×(1-2) **t=6** | | | | |
| 2 | 99.158 | 3.7 | ― | 102.859 | 1642.67<br>Z=1745.529 |
| PD | HVAC: 1×(3-5) **t=1**, 1×(3-5) **t=5**<br>HVDC: 1×(4-6), 1×(2-6) & 1×(5-7) **t=1**, 1×(2-6) **t=2**<br>ES: Bus 7: 51.07 MW, 121.49 MWh **t=1** … 51.07 MW, 121.49 MWh **t=7**<br>*CPU Time: 773 Sec.* | | | | |
| CLL | 100.333 | 5.714 | ― | 106.047 | 1691.127<br>Z=1797.174 |
| PD | HVAC: 1×(3-5) **t=1**, 1×(3-5) **t=4**<br>HVDC: 2×(2-6), 1×(4-6) & 1×(5-7) **t=1**<br>ES: Bus 5: 19.304 MW, 21.449 MWh **t=7**<br>Bus 7: 65.66 MW, 147.75 MWh **t=1** … 65.66 MW, 147.75 MWh **t=7** | | | | |
| 3 | 91.665 | ― | 9.118 | 100.784 | 1647.25<br>Z=1748.033 |
| PD | HVAC: 1×(2-6) **t=1**, 1×(3-5) **t=4**,1×(3-5) **t=7**<br>HVDC: 1×(2-6), 1×(4-6) & 1×(5-7) **t=1**<br>Conversion: (3-5) **t=1**, (2-3) **t=5**<br>*CPU Time: 669 Sec.* | | | | |
| CLL | 99.339 | ― | 5.0027 | 104.342 | 1693.852<br>Z=1798.194 |
| PD | HVAC: 1×(3-5) **t=3**,1×(3-5) **t=5**<br>HVDC: 2×(2-6), 1×(4-6) & 1×(5-7) **t=1**<br>Conversion: (3-5) **t=1** | | | | |
| 4 | 79.664 | 10.432 | 5.002 | 95.1 | 1644.638<br>Z=1739.737 |
| PD | HVAC:1×(2-6) **t=1**<br>HVDC: 1×(2-6), 1×(4-6) & 1×(5-7) **t=1**<br>Conversion: (3-5) **t=1**<br>ES: Bus 7: 80.642 MW, 161.23 MWh **t=1** … 80.642 MW, 161.23 MWh **t=7**<br>Bus 5: 0.219 MW, 0.243 MWh **t=5**, 68.225 MW, 75.806 MWh **t=6,** 139.632 MW, 155.147 MWh **t=7**<br>*CPU Time: 1458 Sec.* | | | | |
| CLL | 92.761 | 6.654 | 5.002 | 104.418 | 1691.707<br>Z=1796.125 |
| PD | HVAC:1×(3-5) **t=4**<br>HVDC: 2×(2-6), 1×(4-6) & 1×(5-7) **t=1**<br>Conversion: (3-5) **t=1**<br>ES: Bus 7: 65.661 MW, 147.75 MWh **t=1** … 65.661 MW, 147.75 MWh **t=7**<br>Bus 5: 37.72 MW, 41.911 MWh **t=7** | | | | |

*PD=Planning Decision  **TPC: Total Planning Cost  ***CLL: Considering Line Losses

Also, one AC to DC conversion is obtained for the transmission line in corridor 3-5, and two ES devices are installed in buses 5 and 7, as given in Table I. The results of ESs installation obtained by scheme 4, in each year of the planning horizon, are illustrated in Fig. 6. According to Table I and Fig. 5, it can be seen that the wind farm in Bus 7 is connected to the primary grid by an HVDC line, which verifies the role of HVDC lines in renewable integration. By considering the CLL cases in Table I, it can be seen, that the total planning cost with considering line losses is increased a little (due to the cost of power losses during the planning horizon), but the decision investments with and without considering line losses do not change significantly. For other test systems, the line losses are ignored, while the VSC losses are considered.

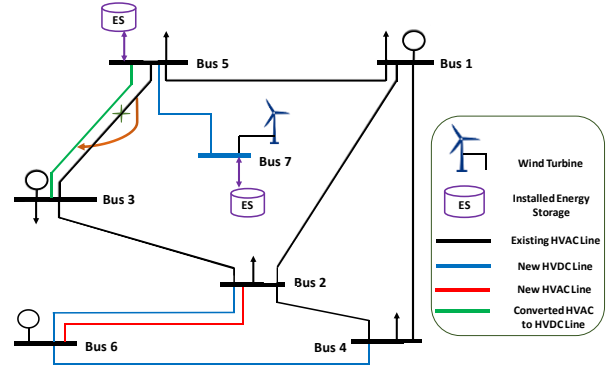

Fig. 5. Results of the fourth scheme over the Garver's 6-bus system

### 2) IEEE 24-bus Test System

As shown in Fig. 7, the modified IEEE 24-bus test system has 34 existing single circuit and 4 double circuit HVAC lines. One existing wind farm is assumed in bus 7, and the wind farms in buses 25 and 26 can be connected to the system by three HVAC and three HVDC candidate lines. The capacity of all wind farms is 800 MW. The wind and load scenarios are handled via the clustering technique, as discussed in the previous section. Also, nine HVAC and four HVDC candidate lines and five candidate buses for the installation of ES devices are considered. In this system, seven existing AC single-circuit lines and two existing AC double-circuit lines are selected for conversion to HVDC. For double circuit lines, it is possible to convert one or both circuits. The complete parameters of all existing and candidate lines and also generating units can be found in [33]. Similar to the 6-bus test system, 4 schemes are considered to discuss the planning results. The complete results of planning for this system are presented in Table II, and for clarification, the obtained results in scheme 4 are depicted in Fig. 7. For this test system, the obtained results confirm that the scheme 4, is more economical than other schemes. In this scheme, the total planning cost is 4.04 10⁶$, 77.246 10⁶$ and 145.656 10⁶$ less expensive than schemes 3, 2 and 1, respectively. According to scheme 2, using only the ES devices results in saving around 68.41 10⁶$ in total planning cost with respect to Scheme 1. By comparing Scheme 3 and Scheme 2, it is concluded that using only AC/DC conversion results in 73.206 10⁶$ cost saving concerning Scheme 2 which confirms the crucial role of HVAC to HVDC conversion in TEP model. According to the presented planning decisions given in Table II, the fourth scheme results in two new HVAC lines in corridor 14-16 and one in corridor 13-14, all constructed in the first year of planning. Also, five HVDC lines including two lines in corridor 7-8, and one line in corridors 2-6, 21-25 and 19-26 are constructed at the first year of planning which confirms the importance of HVDC lines in connecting remote renewable generation to the power system.

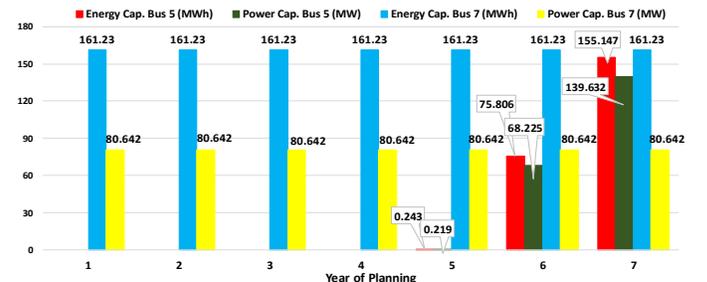

Fig. 6. Power capacity of ESs in scheme 4 over the Garver's 6-bus system

TABLE II
The complete planning results of the IEEE 24-bus test system

| Scheme | TIC (10⁶$) | | | | TOC (10⁶$) |
|---|---|---|---|---|---|
| | HVAC/DC Lines | ESs | Conversion | Total | 8847.532 |
| 1 | 365.647 | ___ | ___ | 365.647 | **TPC (10⁶$):** Z=9213.485 |
| PD | **HVAC:** 2×(14-16), 1×(15-21) & 2×(6-7) **t=1**, 1×(1-3) **t=4**, 1×(7-8) **t=5**, 1×(1-3) **t=6**, 1×(3-9) **t=7** <br> **HVDC:** 2×(7-8), 1×(2-6), 1×(21-25) & 1×(19-26) **t=1**, 1×(19-26) **t=5** <br> *CPU Time: 1603 Sec.* | | | | |
| 2 | 268.79 | 18.689 | ___ | 287.785 | 8857.29 Z=9145.075 |
| PD | **HVAC:** 2×(14-16) & 1×(6-7) **t=1** <br> **HVDC:** 2×(7-8), 1×(19-26) & 1×(21-25) **t=1** <br> **ES: Bus 7:** 25.498 MW, 148.94 MWh **t=1**, 28.114 MW, 235.595 MWh **t=2** … 28.114 MW, 235.595 MWh **t=7** <br> **Bus 10:** 41.536 MW, 532.667 MWh **t=2**, 82.047 MW, 792.122 MWh **t=3**, 92.689 MW, 1022.268 MWh **t=4**, 92.689 MW, 1051.872 MWh **t=5** … 102.7 MW, 1051.872 MWh **t=7** <br> **Bus 26:** 37.824 MW, 81.872 MWh **t=3**, 66.799 MW, 180.415 MWh **t=4**, 66.799 MW, 415.128 MWh **t=5**, 66.799 MW, 1202.685 MWh **t=6** … 66.799 MW, 1202.685 MWh **t=7** <br> *CPU Time: 4715 Sec.* | | | | |
| 3 | 202.505 | ___ | Full: 27.227 Half: 41.474 | 271.207 | 8800.662 Z=9071.869 |
| PD | **HVAC:** 2×(14-16) & 1×(6-7) **t=1** <br> **HVDC:** 2×(7-8), 1×(2-6), 1×(21-25) & 1×(19-26) **t=1** <br> **Full Conversion:** (8-10) **t=1**, (1-3) **t=3**, (7-8) **t=4**, **Half Conversion:** (15-21) & (19-20) **t=1** <br> *CPU Time: 4092 Sec.* | | | | |
| 4 | 202.505 | 6.739 | Full: 26.738 Half: 41.474 | 277.458 | 8790.371 Z=9067.829 |
| PD | **HVAC:** 2×(14-16) & 1×(13-14) **t=1** <br> **HVDC:** 2×(7-8), 1×(2-6), 1×(21-25) & 1×(19-26) **t=1** <br> **Full Conversion:** (8-10) **t=1**, (1-3) & (7-8) **t=4**, **Half Conversion:** (15-21) & (19-20) **t=1** <br> **ES: Bus 26:** 62.473 MW, 104.056 MWh **t=3**, 62.473 MW, 168.734 MWh **t=4**, 62.473 MW, 407.753 MWh **t=5**, 62.473 MW, 1175.429 MWh **t=6**, 62.473 MW, 1175.429 MWh **t=7** <br> *CPU Time: 8895 Sec.* | | | | |

Besides, three AC to DC full conversion and two half conversion (i.e. one circuit of existing double circuit HVAC line) are obtained for transmission lines in corridors 8-10, 1-3, 7-8, 15-21 and 19-20, respectively. One ES is installed in bus 26 as given in Table II. To illustrate the impacts of daily cycles of ESs in the presence of renewable resources, Fig. 8, is presented. Based on Fig. 8, the power generation of conventional and renewable units, load and the level of energy of ESs are presented for all 96 hours of last year of planning. The charging/discharging of ESs is consistent with wind and load variations which facilitate the utilization of renewable generation, reducing generation cost, and economic planning decisions.

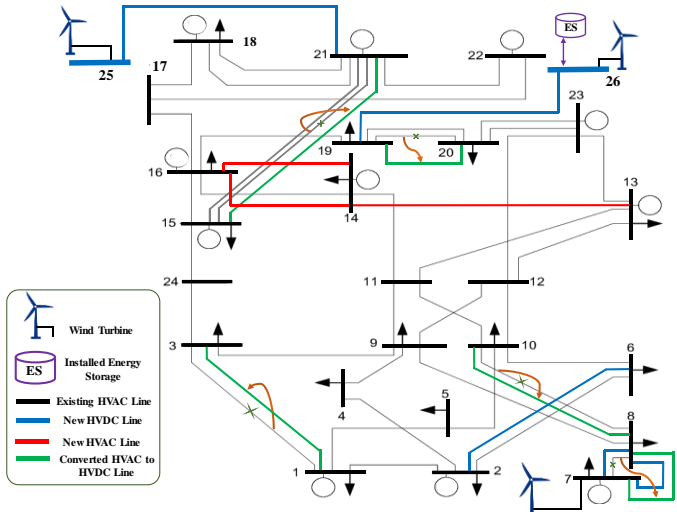

Fig. 7. Results of the fourth scheme over the IEEE 24-bus test system

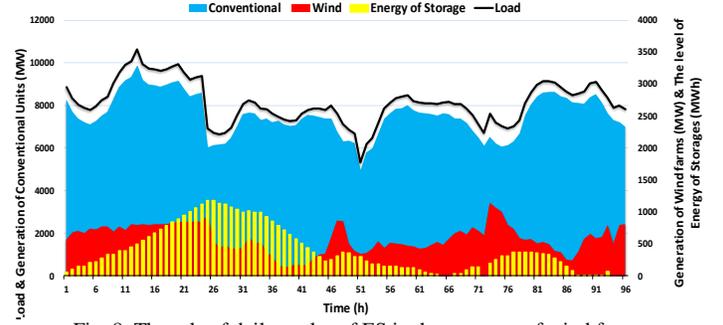

Fig. 8. The role of daily cycles of ES in the presence of wind farms

The amount of flexible ramp and spinning reserves impact the planning results. A total amount of 5% of total renewable generation is assigned in each hour as the flexible ramp reserve for compensating the renewable uncertainties as given in (9a) to (9c). In this regard, by increasing the amount of required reserves, the total cost of transmission expansion is increased. For justification of this fact, a new case is defined for the scheme 4. In Case 1, the flexible ramp reserve and spinning reserve are increased to 12% of total renewable generation, and also 8% of total load in each hour, respectively. The obtained results are given in Table III. Besides, the number of input scenarios or representative days impact the planning results. Therefore, a new case named by Case 2 is defined for Scheme 4 to justify the utilized scenario-based method for handling uncertainties. In Case 2, the number of considered representative days is decreased to two days (i.e. 2×24=48 hours). The obtained results are reported in Table III. It can be seen that by increasing the amount of required flexible and spinning reserves (i.e. Case 1), the total planning cost is 66.75 10⁶$ more than the original Scheme 4, which shows the importance of required reserves. Also, in Case 2, it can be seen that the total planning cost is increased to 9741.322 10⁶$ which is significantly higher than the original total planning cost in Scheme 4 with four representative days (i.e. 4×24=96 hours).

*3) IEEE 118-bus Test System*
The modified IEEE 118-bus test system has 172 existing single circuit and 7 double circuit HVAC lines. Also, two buses (i.e. buses 119 and 120), can be connected to the system by two candidate HVAC and two candidate HVDC lines. The capacity of all wind farms is 800 MW. Also, eighteen HVAC and eight HVDC candidate lines and eight candidate buses for the installation of ES devices are considered. In this system, eight existing AC single-circuit lines and four existing AC double-circuit lines are selected for conversion to HVDC. The complete parameters of all existing and candidate lines and also generating units can be found in [33]. Similar to previous test systems, 4 schemes are considered to discuss the planning results. The complete results of planning decisions are presented in Table IV. The obtained results for this test system confirm that the scheme 4, is more economical than other schemes. In this scheme, the total planning cost is 22.816 10$^6$\$, 38.397 10$^6$\$ and 60.301 10$^6$\$ less expensive than schemes 3, 2 and 1, respectively. By comparing Scheme 3 and Scheme 2, it is concluded that using only AC/DC conversion results in 15.581 10$^6$\$ cost saving concerning Scheme 2. It is noted that in Scheme 2, only the ES installation is considered without AC/DC conversion. According to the presented planning decisions given in Table IV, the fourth scheme results in one new HVAC line in corridors 17-18, 77-78 and 110-111, all constructed in the first year of planning. In the second and fifth years of planning, a new HVAC line is constructed in corridor 110-112 and 69-77, respectively. Also, four HVDC lines in corridors 5-6, 69-77, 59-119 and 116-120 are constructed in the first year of planning which confirms the importance of HVDC lines in connecting remote renewable generation to the power system. Construction of both HVAC and HVDC lines in corridor 69-77, confirm the importance of hybrid AC/DC TEP. Besides, two AC to DC full conversion are obtained for transmission lines in corridors 23-32, and 42-49. The full converted line in corridor 42-49 is a double circuit line. Two ES devices are installed in buses 119 and 120 as given in Table IV. The computational times of the proposed model are reported in Table I, II and IV, for all test cases under all schemes. Although, in long term planning studies, the computational time is not very critical and instead the optimality of the solution has higher priority, the CPU times of the proposed model are acceptable.

TABLE III
The planning results for Scheme 4 of the IEEE 24-bus test system by changing the required reserve (i.e. Case 1) and number of representative days (i.e. Case 2)

| | TIC (10⁶$) | | | | TOC (10⁶$) |
|---|---|---|---|---|---|
| | HVAC/DC Lines | ESs | Conversion | Total | |
| **Scheme 4** | 202.505 | 6.739 | Full: 26.738 Half: 41.474 | 277.458 | 8790.371 **TPC (10⁶$):** Z= 9067.829 |
| **Case 1** | 202.505 | 6.739 | Full: 26.738 Half: 41.474 | 277.458 | 8857.121 Z=9134.579 |
| **Case 2** | 202.505 | 11.353 | Full: 11.318 Half: 20.835 | 246.012 | 9495.31 Z=9741.322 |

TABLE IV
The complete planning results of the IEEE 118-bus test system

| Scheme | TIC (10$^6$\$) | | | | TOC (10$^6$\$) |
|---|---|---|---|---|---|
| | HVAC/DC Lines | ESs | Conversion | Total | 11320.9 |
| 1 | 198.693 | — | — | 198.693 | TPC (10$^6$\$): Z=11519.593 |
| PD | \multicolumn{5}{l}{**HVAC:** 1×(17-18), 1×(49-51), 1×(77-78) & 1×(110-111) **t=1**, 1×(37-39) & 1×(110-112) **t=2**, 1×(69-77) **t=5**, 1×(77-78) **t=6**} | |
| | \multicolumn{5}{l}{**HVDC:** 1×(5-6), 1×(23-32), 1×(69-77), 1×(59-119) & 1×(116-120) **t=1**} | |
| | \multicolumn{5}{l}{*CPU Time: 2194 Sec.*} | |
| 2 | 196.635 | 26.154 | — | 222.789 | 11274.9 Z=11497.689 |
| PD | \multicolumn{5}{l}{**HVAC:** 1×(17-18), 1×(77-78) & 1×(110-111) **t=1**, 1×(110-112) **t=2**, 1×(37-39) & 1×(49-51) **t=3**, 1×(69-77) **t=5**, 1×(77-78) **t=6**} | |
| | \multicolumn{5}{l}{**HVDC:** 1×(5-6), 1×(69-77), 1×(59-119) & 1×(116-120) **t=1**, 1×(23-32) **t=2**} | |
| | \multicolumn{5}{l}{**ES: Bus 119**: 145.015 MW, 1928.234 MWh **t=1**, 145.015 MW, 1998.343 MWh **t=2** ... 145.015 MW, 1998.343 MWh **t=7**} | |
| | \multicolumn{5}{l}{**Bus 120:** 110.432 MW, 974.392 MWh **t=1** ... 110.432 MW, 974.392 MWh **t=7**} | |
| | \multicolumn{5}{l}{*CPU Time: 6367 Sec.*} | |
| 3 | 166.252 | — | Full: 16.253 Half: 5.507 | 188.012 | 11294.096 Z=11482.108 |
| PD | \multicolumn{5}{l}{**HVAC:** 1×(17-18), 1×(77-78) & 1×(110-111) **t=1**, 1×(110-112) **t=2**, 1×(69-77) **t=5**, 1×(77-78) **t=6**} | |
| | \multicolumn{5}{l}{**HVDC:** 1×(5-6), 1×(69-77), 1×(59-119) & 1×(116-120) **t=1**} | |
| | \multicolumn{5}{l}{**Full Conversion:** (42-49) **t=2**, (23-32) **t=3**, **Half Conversion:** (49-54) **t=1**} | |
| | \multicolumn{5}{l}{*CPU Time: 5561 Sec.*} | |
| 4 | 166.111 | 29.132 | Full: 16.009 Half: 0 | 211.252 | 11248.04 Z=11459.292 |
| PD | \multicolumn{5}{l}{**HVAC:** 1×(17-18), 1×(77-78) & 1×(110-111) **t=1**, 1×(110-112) **t=2**, 1×(69-77) **t=5**} | |
| | \multicolumn{5}{l}{**HVDC:** 1×(5-6), 1×(69-77), 1×(59-119) & 1×(116-120) **t=1**} | |
| | \multicolumn{5}{l}{**Full Conversion:** (23-32) **t=2**, (42-49) **t=3**} | |
| | \multicolumn{5}{l}{**ES: Bus 119:** 184.278 MW, 1914.482 MWh **t=1** ... 184.278 MW, 1914.482 MWh **t=7**} | |
| | \multicolumn{5}{l}{**Bus 120:** 132.063, 756.881 MWh **t=1** ... 132.063 MW, 756.881 MWh **t=7**} | |
| | \multicolumn{5}{l}{*CPU Time: 12204 Sec.*} | |

## V. Conclusion

In this paper, a dynamic hybrid AC/DC transmission expansion planning problem was proposed. Different planning decisions including individual HVAC, individual HVDC, HVAC/HVDC conversion, and Energy Storage devices were considered to optimize the transmission planning. The major findings of this paper are summarized as follows. 1) The HVDC options play an important role in increasing the transfer capability of the power system, especially under the high penetration of remote renewable resources. 2) Along with constructing new HVDC lines, the HVAC to HVDC line conversion seems to be a proper decision planning that results in a further significant reduction of planning cost. 3) Under the high penetration of renewable resources, the ES devices result in better transmission planning configuration due to its capability in charging and discharging in daily cycles. For IEEE 24-bus and 118-bus test systems, it was shown that the HVAC to HVDC line conversion is more economical than the ES option. 4) To capture the realistic variations of load and wind power uncertainties, some representative days should be extracted for the hybrid AC/DC TEP model. Otherwise, the daily cycles of ES devices in the presence of renewable resources are ignored, which results in impractical results.

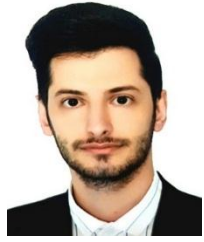

**Mojtaba Moradi-Sepahvand** received his BSc and MSc degrees in Power System Engineering from Lorestan University, Khorram-Abad, and Shahid Chamran University of Ahvaz, Ahvaz, Iran, in 2015 and 2017, respectively. His research interests include modern power system planning and operation, renewable integration, and power system observability. He is currently pursuing the Ph.D. degree in Power System Engineering at K.N. Toosi University of Technology, Tehran, Iran.

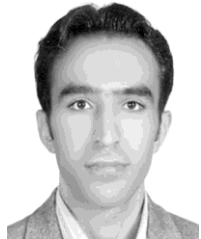

**Turaj Amraee** (SM'16) received the Ph.D. degree in power system engineering from Sharif University of Technology, Tehran, Iran in collaboration with Grenoble-INP University, Grenoble, France, in 2010. He is currently an Associate Professor with the Electrical Engineering Department, K.N. Toosi University of Technology, Tehran, Iran. He is also a consultant with the Iran Grid Management Company in the areas of national grid monitoring and security assessment using Wide Area Measurement System. His research interests include short term operation and long-term planning of modern power systems, renewable integration and operation, security assessment and stability issues in smart grids, data mining applications in power systems.